\documentclass{article}

\usepackage{graphicx}

\newtheorem{defn}{Definition}

\newtheorem{teor}[defn]{Theorem}

\newtheorem{lema}[defn]{Lemma}
\newtheorem{rema}[defn]{Remark}

\newtheorem{prop}[defn]{Proposition}
\newtheorem{defi}[defn]{Definition}

\makeatother

\font\ddpp=msbm10  at 11 truept 
\def\R{\hbox{\ddpp R}}     
     
\def\L{\hbox{\ddpp L}}    

\def\Z{\hbox{\ddpp Z}}
     
\def\N{\hbox{\ddpp N}}

\newcommand{\Tau}{{\cal T}}
\newcommand{\B}{{\cal B}}
\newcommand{\bb}{\overline{{\cal B}}}
\newcommand{\noi}{\noindent}

\newcommand{\be}{\begin{equation}}
\newcommand{\ee}{\end{equation}}
\newcommand{\cvd}{{\rule{0.5em}{0.5em}}\smallskip}

\newcommand{\ben}{\begin{enumerate}}
\newcommand{\een}{\end{enumerate}}

\newcommand{\bit}{\begin{itemize}}
\newcommand{\eit}{\end{itemize}}
\newcommand{\edoc}{\end{document}}

\begin{document}

\title{Causal hierarchy of spacetimes, temporal functions and smoothness of Geroch's splitting. \\ A revision }
\author{ Miguel S\'anchez
\\ Depto. Geometr\'{\i}a y Topolog\'{\i}a, Universidad de Granada, \\ Facultad de Ciencias, Avda. Fuentenueva s/n, E-18071 Granada, Spain}%
\maketitle

\begin{abstract}

\noindent After the heroic epoch of Causality Theory, problems concerning the smoothability of time functions and Cauchy hypersurfaces remained as unanswered folk questions. Just recently  solved, our aim is to discuss the state of the art on this topic, including self-contained proofs  for  questions on causally continuous, stably causal  and globally hyperbolic spacetimes.

\end{abstract}


\smallskip


\section{Introduction}

The following three highlights in the roots of the theory of Causality, become relevant to understand the causal hierarchy of spacetimes: (a) Geroch's splitting theorem \cite{Ge}, which ensures the existence of a Cauchy time function (and, then a topological splitting) in any globally hyperbolic spacetime, (b) the introduction by Hawking and Sachs \cite{HS} of causal continuity, as an intermediate level in the ladder of causality, fulfilled  when the future and past volume functions (essential in Geroch's proof) become time functions, and (c) Hawking's proof of the existence of a time function in any stably causal spacetime \cite{Ha}, obtained by averaging volume functions (not necessarily continuous) of metrics close to the original one.

Here, concepts such as Cauchy hypersurfaces, Geroch's splitting or time functions appear at a topological level, and the possibility to smooth them remained as open folk questions. Briefly (see \cite[Section 2]{BSc} for an expanded summary), a smoothing procedure claimed by Seifert \cite{SeA} (cited in \cite{HE} and then in many references) presented some gaps. Thus, Sachs and Wu \cite[p. 1155]{SW} raised the issue of the existence of a {\em smooth} Cauchy hypersurface in any globally hyperbolic spacetime. In spite of Dieckmann's progress (\cite{Di}, see Section \ref{s31}), his attempt \cite{DiA} was not sufficient. Thus, as pointed out  in Beem and Ehrlich's book (including the last edition, with Easley \cite{BEE}) the problem has persisted. Recently, we have given a full solution by proving not only the existence of a {\em smooth and spacelike} Cauchy hypersurface \cite{BS1}, but also \cite{BS2}: (i) the existence of a Cauchy temporal function (and, then, an orthogonal splitting) in any globally hyperbolic spacetime, and (ii) the existence of a temporal function whenever a time function exists. Here, ``temporal'' means not only ``smooth and time'' function but also with timelike gradient everywhere (see Section \ref{s2} for definitions). This subtlety becomes important in order to ensure that the levels of the function are spacelike and, specially, that Geroch's topological splitting not only can be smoothed but also strengthened in an orthogonal splitting. 

In the present article, our aim is to revise the state of art on this topic, including a full proof of the orthogonal splitting of globally hyperbolic spacetimes and the equivalence between stable causality and the existence of a temporal function. The original proofs are rewritten in a self-contained way, even though the smoothing procedure, which is quite technical and long, is only sketched  (we refer to \cite{BSc} for a expanded version). Concretely:

In Section \ref{s2}, basic notation and background are introduced. 

In Section \ref{s3}, first we discuss the problems related to the  measures which are admissible in order to define future and past volume functions $t^-, t^+$. In Subsection \ref{s32} we focus on the equivalence between the  property of being  distinguishing  and the increasing of the volume functions on any inextendible causal curve, Proposition \ref{voldis}. Then, in Subsection \ref{s33} we show  the equivalence between causal continuity and the continuity of $t^-, t^+$, Theorem \ref{caucon}. The proofs are direct, and intermediate relations with reflectivity are avoided (compare with \cite{Di}, \cite[pp. 68-71]{BEE}). Finally, in Subsection \ref{s34}, causally simple spacetimes (and, then, globally hyperbolic ones) are shown to be causally continuous. 

In Section \ref{s5}, stable causality is revisited. First, we recall that any stably causal spacetime admits a time function, and that any spacetime admitting a  temporal function is stably causal. Therefore, the equivalence between stable causality and the existence of a time function will be completed by showing that, if a time function exists, then a temporal function will exist (see Fig. \ref{fig2} and Remark \ref{reojo}). Nevertheless, the proof of this fact is not straightforward, and it is postponed to the smoothing procedure, Subsection \ref{s63}.

In Section \ref{s4},  Geroch's topological splitting theorem is proved. The reader can obtain a complete self-contained proof of this theorem by taking into account, in addition to this section: (1) the existence of an admissible measure $m$ (Subsection \ref{s31}),  (2)  volume functions $t^\pm$ are time functions in causally continuous spacetimes (Theorem \ref{caucon}, proved from Proposition \ref{voldis}(b) and Lemma \ref{l33a}) and  (3) any globally hyperbolic spacetime is causally continuous (Remark \ref{r411}(1)).

In Section \ref{s6} the smoothing procedure is briefly sketched. The main goal is to find a Cauchy temporal function in any globally hyperbolic spacetime (Subsection \ref{s62}); a simplification of the arguments yields a temporal function whenever a time function exists (Subsection \ref{s63}). As a previous result -with interest in its own right- the existence of a spacelike Cauchy hypersurface is proved, Section \ref{s61}. Finally, in Section \ref{s7} the  results are discussed and new open questions are posed.

\section{Preliminaries} \label{s2}
$(M,g)$ will denote  a spacetime, i.e., a connected time-oriented $C^k$ Lorentzian $n_0-$manifold, $n_0\geq 2, k=1,2,... \infty$; {\em smooth} will mean $C^k$-differentiable. The signature will be chosen $(-,+,\dots, +)$ and, thus, for a timelike (resp. causal, lightlike, spacelike) tangent vector $v\neq 0$, one has $g(v,v)<0$ (resp. $\leq 0, =0, >0$); following \cite{O}, vector 0 will be regarded as spacelike. Timelike and causal curves are defined consistently, and are always assumed piecewise smooth (with the two limit tangent vectors at the break in the same causal cone). The chronological and causal future and past of any $p\in M$ are denoted as standard $I^+(p), I^-(p), J^+(p), J^-(p)$ and, in general, the assumed geometrical and notational background  can be found in
well--known books as \cite{BEE,HE,O,Pe}; for example, $I^+(p)=\{q\in M: p<<q\}, J^-(p)=\{q\in M: q\leq p\}$. If $A \subset M$, then $\bar A, \partial A$ denote, resp., the  closure and 
frontier of $A$. 
Hypersurfaces  will be always embedded without boundary (thus, ``closed hypersurface  $M$'' means closed as a subset of $M$). In general, they are only topological, but {\em spacelike} hypersurfaces will be regarded as smooth.

It is well-known the causal hierarchy of spacetimes \cite[p. 73]{BEE}, \cite{HS}:

\begin{center}
Globally hyperbolic  $\Rightarrow$ Causally simple $\Rightarrow$ Causally continuous \\ $\Rightarrow$ Stably causal
$\Rightarrow$ Strongly causal \\ $\Rightarrow$ Distinguishing $\Rightarrow$ Causal $\Rightarrow$ Chronological
\end{center}
Recall that $(M,g)$  is {\em chronological} (resp. {\em causal}) if it does not contain closed timelike (resp. causal) curves. The spacetime is {\em distinguihing} if it is past distinguishing ($p\neq q  \Rightarrow I^-(p)\neq I^-(q)$, i.e., the set valued function $I^-$ is injective) and  future  distinguishing (defined analogously). In particular, for any future-
directed causal curve $\gamma$, if $t<t'$ then 
$I^-(\gamma(t))$ (resp. $I^+(\gamma(t'))$) is  included strictly in $I^-(\gamma(t'))$ (resp. $I^+(\gamma(t))$). Roughly,  $(M,g)$ is strongly causal if it does not contain ``almost closed'' causal curves,  and stably causal if, after opening slightly the light cones, the spacetime remains causal (see Definition \ref{dsc}). All these definitions, which correspond to the second and third lines in the diagram above, are rather intuitive. In fact, it is straightforward  not only to check the implications, but also to find examples which show that  no converse implication hold.

The definition of {\em causal continuity} is somewhat more involved: it requires  the outer continuity of the set-valued functions $I^-, I^+$, plus to be distinguishing. Function, say, $I^-$, is called inner (resp. outer) continuous at some $p\in M$ if, for any compact subset 
$K\subset I^-(p)$ (resp. $K\subset M\backslash \overline{ I^-(p) }$),  there exists an open neighborhood $U \ni p$ such that $K \subset I^-(q)$ 
(resp. $K\subset M\backslash \overline{I^-(q)}$) for all $q \in U$. Functions $I^\pm$ are always inner continuous \cite[Sect. 1.6]{HS}, but it is easy to construct examples non-outer continuous  (see for example Fig. \ref{fig1}).


Nevertheless,  outer continuity itself does not imply a good causal behavior: functions 
$I^\pm$ are outer continuous in any totally vicious spacetime, i.e., those  (obviously,  non-chronological) spacetimes such that 
$I^\pm(p)=M$ for all $p\in M$.  This is the reason to add being distinguishing in the definition   of causal continuity. Summing up: 
{\begin{quote}
A spacetime is called causally continuous if the set valued functions $I^-, I^+$ are both, continuous and injective.
\end{quote} Recall that the  stable causality of a causally continuous spacetime is not obvious (see Section \ref{s5}). There are alternative definitions of causal continuity \cite[Theorem 2.1]{HS}; in particular,  a spacetime is causally continuous if and only if it is distinguishing and reflecting (i.e., past reflecting $I^+(p) \supseteq I^+(q) \Rightarrow I^-(p) \subseteq I^-(q)$ and future reflecting
$I^-(p) \subseteq I^-(q) \Rightarrow I^+(p) \supseteq I^+(q)$).

Globally hyperbolic spacetimes can be defined as the strongly causal ones with compact diamonds $J^+(p)\cap J^-(q)$ for any $p, q$ (this definition is somewhat different, but equivalent 
 to Leray's original one, in terms of the compactness of the space of causal curves connecting $p$ and $q$ \cite{Av, Se}). They were characterized by Geroch as those possesing a Cauchy hypersurface, that is, an achronal subset $S$ which is crossed exactly once by any inextendible timelike curve. Remarkably, such a $S$ must be a (topological) hypersurface, and it is intersected by any inextendible causal curve.  It is easy to check that any globally hyperbolic spacetime must be {\em causally simple}, i.e., distinguishing with closed $J^\pm(p)$ for any
 $p\in M$, see \cite[p.65]{BEE}. And, as we will see in Subsection \ref{s34}, any causally simple  spacetime is causally continuous, which becomes essential for Geroch's proof.

Finally, the following concepts will be closely related to stable causality and global hyperbolicity.

\begin{defi} {\em 
(1) A function  $t: M \rightarrow \R$ is a {\em time function} if it is continuous and strictly increasing  on any future-directed causal curve. If, additionally, each level hypersurface $S_a=t^{-1}(a)$ is a Cauchy hypersurface (for all $a$ in the image Im $t$), then $t$ is a {\em  Cauchy time function}.

(2) A smooth function  $\Tau: M \rightarrow \R$ is a {\em  temporal function} if its gradient is everywhere timelike and past-pointing. If, additionally, each (spacelike) level  hypersurface $S_a=\Tau^{-1}(a)$ is a Cauchy hypersurface (for all $a$ in Im$\Tau$), then $\Tau$ is a {\em  Cauchy temporal function}.
}
\end{defi}
The image of such $t$ or $\Tau$ is always an open interval $I$ of $\R$; composing with an increasing diffeomorphism  $I\rightarrow \R$, an onto time or temporal function can be constructed.

\begin{rema}\label{r22} {\em A temporal function is always a time function, but even a smooth time function may be a non-temporal one. In general, a Cauchy hypersurface maybe intersected by a causal curve in more than a point (say, a segment), but this is not the case of the level hypersurfaces of a (Cauchy) time function, which are acausal.}
\end{rema}

\section{Volume functions and causal continuity} \label{s3}

\subsection{Admissible measures}\label{s31}

Geroch's proof depends on a function constructed from the volumes of $I^\pm(p), p\in M$. 
But for this purpose, such volumes must be finite and, thus, the natural measure associated to the metric may not be useful. Nevertheless, there is a straightforward method to modify such measure and obtain a new {\em finite} measure which preserves the good properties (for the differentiable and causal structures of $(M,g)$) of the original one. 

Concretely, consider the following construction of a Borel measure on $M$ (see \cite[p. 199]{HE}, \cite[p. 67]{BEE}), that is, a measure on the $\sigma-$algebra generated by the open subsets of $M$. Without loss of generality, we will assume that $M$ is orientable because, otherwise, we can reason with the orientable Lorentzian double-covering $\Pi: \tilde M \rightarrow M$, and define the measure of any Borelian $A\subset M$ as (one half of) the measure of $\Pi^{-1}(A)$. Choose an orientation, and let $\omega$ be the oriented volume element  associated to the metric $g$. Fix any  covering of $M$  by open subsets with  $\omega$-measure smaller than 1, and take a partition of the unity $\{\rho_n\}_{n\in \N}$ subordinated to the covering. Now, define the measure $m$ as the one associated to the volume element
\be \label{eomegas}
\omega^*= \sum_{n=1}^\infty 2^{-n} \rho_n \omega .
\ee
Notice that, chosen any auxiliary Riemannian metric $g_R$ with associated oriented volume element $\omega_R$, necessarily
$$ \omega^*= e^{ w} \omega_R$$
for some smooth function $w$. Thus, $\omega^*$ is also the volume element associated to the conformal Riemannian metric  (which depends on the dimension $n_0$ of $M$) $g_R^* = e^{2w/n_0}g_R$, and $m$ can be regarded as the natural measure associated to $g_R^*$. 

Such measures associated to Riemannian metrics on a manifold are very well-known. We can assume that $m$ is completed in the standard way, by adding to the Borel sigma algebra all the subsets of any subset of measure 0 (which are regarded as new subsets of measure 0).  By Sard's theorem, the subsets of measure 0 are intrinsic to the differentiable structure of $M$. Thus,
a subset $A\subset M$ will have zero-measure if and only if for any (differentiable) chart $\varphi: U \subseteq M \rightarrow \R^n$, the outer Lebesgue measure of $\varphi(U\cap A)$ is 0. Moreover, the whole manifold can be regarded as a starshaped domain, up to a zero measure subset. In fact,  
assuming that $g_R$ has been chosen complete, fix $p_0$ and remove the cut locus Cut$(p_0)$; then the exponential map exp$_{p_0}:  D\subseteq T_{p_0}M \rightarrow M\backslash $Cut$(p_0)$ is a diffeomorphism, where $D$ is the maximal starshaped domain. By choosing an orthonormal basis on $T_{p_0}M$, and then identifying $T_{p_0}M$ with $\R^n$, one has an almost everywhere chart
$\varphi_0:  M\backslash \hbox{Cut}(p_0)\rightarrow \tilde D\subseteq \R^n$. 
Roughly, this allows to transplantate the integration of functions on $M$ with respect to $\omega^*$,  to the usual Lebesgue integration on $\R^n$: a function $f$ on $M$ will be integrable if and only if $|g_R^*| \cdot f\circ \varphi_0^{-1}$ is Lebesgue integrable on $\tilde D\subseteq \R^n$, where $|g_R^*|$ is the square root of the determinant of the matrix $(g_R^*)_{ij}$ in the coordinates given by $ \varphi_0$.
In particular, Lebesgue's theorems of monotonous and dominated convergence hold, the measure $m$ can be recovered as the integration of the characteristic function of the measurable subsets of $M$, and $m$ is a regular measure in the sense below.

The relevant properties of the so--defined measure $m$ will be the following ones (the second and the fourth hold obviously, because they are satisfied by the usual Lebesgue measure on $\R^n$):
\ben
\item Finiteness: $m(M)<\infty$.

This is straightforward from (\ref{eomegas}) and one can
 normalize $m(M)=1$.
\item For any non-empty open subset $U$, $m(U)>0$.

\item The boundaries $\partial I^+(p), \partial I^-(p)$ have measure 0, for any $p\in M$.

This holds for $m$ because  
$\partial I^+ (p)$, $\partial I^- (p)$ are closed, embedded, achronal hypersurfaces  \cite[Proposition 6.3.1]{HE}; thus, for any (differentiable) chart, they can be written as Lipschizian graphs, which have  0 measure.

\item Regularity: for any measurable subset $A\subseteq M$ there exists a sequence $\{G_n\}$ of open subsets which contains $A$, and a sequence $\{K_n\}$ of compact subsets contained in $A$ such that $G_n \supset  G_{n+1}$, $K_n \subset  K_{n+1}$ for all $n$ and:
$$ m(A) = \lim_n m(G_n) = \lim_n m(K_n).$$
\een
These properties had been used implicitly, even in papers where the measure $m$ had not been constructed as above. Dieckmann \cite{Di} (see also \cite[Sect. 3.2]{BEE}) stated explicitly the necessity of the three first ones, and 
emphasized the necessity of the third, which ensures 
$$m(I^+(p))=m(\overline{I^+(p)})= m(J^+(p))$$ for all $p$, and analogously for $I^-$. Obviously, the third property cannot be deduced from the first an the second ones. In fact, consider a measure $m$ as above, choose a point $q\in M$, and construct a new measure $m'$ regarding $q$ as an 
{\em atom}\footnote{that is, a measurable subset $C$ with positive measure $m'(C)$ and which contains no measurable subset $B\subset C$ with $0<m'(B)<m(C)$.}, say: $m'(A)=m(A)+1$ if $q\in A$, $m'(A)=m(A)$ if $q\not\in A$, for all measurable subset $A$. Clearly, if $q\in J^+(p)\backslash I^+(p)$ then  $m'(\partial I^+(p))= 1$.

Measures satisfying the three first properties are called {\em admissible} \cite[Definition 3.19]{BEE}. The fourth one is not  restrictive under our approach. Nevertheless, we will not need this property, but a weaker one which can be deduced for any admissible measure. Concretely, let $U\subseteq M$ be open, and choose a sequence of compact subsets $\{K_n\}$ such that:
\be\label{succom} K_n \subset K_{n+1} \subset U \quad \forall n\in \N, \quad \quad U = \cup_{n\in \N} K_n  \ee
(such a sequence always exists by the paracompactness of $M$). Then,
$
m(U)= \lim_n m(K_n)
$ (see for example \cite[Teor. 2.1.3]{GR}). 
In particular, as the admissible measure is finite, given $\epsilon>0$: 
\be \label{maux}
0\leq  m(U)-m(K_n) < \epsilon , 
\ee
 for large $n$. 

\subsection{Volume functions and generalized time functions}\label{s32}

Now, consider the future $t^-$ and past $t^+$ volume functions associated to $m$, defined as:
$$t^-(p)= m(I^-(p)), \quad  \quad t^+(p)= -m(I^+(p)), \quad  \quad \forall p \in M.$$
Notice that  these functions are non-decreasing on any future-directed causal curve. In fact, the  sign -  is introduced for $t^+$ because of this reason (in what follows, we will reason for $t^-$ and the reasonings for $t^+$ will be analogous). Even more:

\begin{prop}\label{voldis} The spacetime $(M,g)$ is:
\bit
\item[(a)] Chronological if and only if $t^-$ (resp. $t^+$) is strictly increasing on any future-directed timelike curve.

\item[(b)] Past (resp. future) distinguishing if and only if $t^-$ (resp. $t^+$) is strictly increasing on any future-directed causal curve.
\eit
\end{prop}
{\em Proof.} (a) ($\Rightarrow$). If $p << q$ but $t^-(p) = t^-(q)$, necessarily almost all the points (i.e., all but a 0-measure subset) in the  open subset $I^+(p)\cap I^-(q)$ lie in 
$I^-(p)$. Thus, $(I^+(p)\cap I^-(q)) \cap I^-(p)$ is non-empty, and any point $r$ in this intersection satisfies $p << r << p$, that is, $r$ is crossed by a closed timelike curve. ($\Leftarrow$). Obviously, $t^-$ is constant on any closed timelike (or even causal) curve.  

(b) ($\Rightarrow$). Assume that $p\leq q$ and $p\neq q$ but $t^-(p)=t^-(q)$. Then, almost all the points of $I^-(q)$ are included in $I^-(p)$. Choose a sequence $\{q_n\} \subset I^-(p)\cap I^-(q)$ converging to $q$.
Recall that, necessarily then $I^-(q_n) \subset I^-(p)$ for all $n$, and $I^-(q) = \cup_n I^-(q_n)$. 
But this implies $I^-(q) \subset I^-(p)$ and, as the reversed inclusion is obvious, the spacetime is 
non-past distinguishing.
($\Leftarrow$). If $I^-(p) = I^-(q)$ with $p\neq q$, choose a sequence $\{p_n\} \subset I^-(p)$ which converges to $p$, and a sequence of timelike curves $\gamma_n$ from $q$ to $p_n$. By construction, the limit curve $\gamma$ of the sequence\footnote{Alternatively, one can take a quasilimit \cite[Prop. 14.8]{O}.}  \cite[Lemma 14.2]{BEE}  starting at $q$ is a (non-constant) causal curve and $I^-(p)\subseteq I^-(\gamma(t)) \subseteq I^-(q)$ for all $t$. Thus, the equalities in the inclusions hold, and $t^-$ is constant on $\gamma$. $\cvd$

\smallskip

\noindent As suggested in the proof of (a)($\Leftarrow$), causal but non-distinguishing spacetimes cannot be characterized in this way. 

Notice that functions $t^\pm$ are not necessarily continuous in a distinguishing spacetime (Fig. \ref{fig1}) -the discontinuity points can be studied further \cite[Prop. 1.7]{Di}. Nevertheless, Proposition \ref{voldis} suggests to define a {\em generalized time function} as a (non-necessarily continuous) function 
strictly increasing on any future-directed causal curve (i.e., time functions are the continuous generalized time functions). Thus, Proposition \ref{voldis}(b) can be reparaphrased as: \begin{quote} a spacetime is past (resp. future) distinguishing if and only if $t^-$ (resp. $t^+$) is a generalized time function for one (and then for any) admissible measure.\end{quote}

\subsection{Continuous volume functions}\label{s33}

Next, we will see that the  outer continuity of $I^\pm$ are equivalent to the continuity of the past and future volume functions. As the property of being distinguishing has also been characterized in terms of the volume functions, a full characterization of causal continuity in terms of the properties of $t^-, t^+$ will be obtained, Theorem \ref{caucon}. 

The characterization of the continuity of $t^-, t^+$, Proposition \ref{p33a}, is straightforward from the following lemma.
\begin{lema} \label{l33a} 
(a) The inner continuity of $I^-$ (resp. $I^+$) is equivalent to the lower (resp. upper) semi-continuity of $t^-$ (resp. $t^+$). Thus, this holds always. 

(b) The outer continuity of $I^-$ (resp. $I^+$) is equivalent to the upper (resp. lower) semi-continuity of $t^-$ (resp. $t^+$).

\end{lema}
{\em Proof.} We will reason for $I^-$.

\smallskip

\noindent (a) As $I^-$ is always inner continuous, only the implication to the right must be proved. Thus, let $\{p_n\}\rightarrow p$, fix $\epsilon>0$ and let us prove $t^-(p_n) >t(p)-\epsilon$ for large $n$. By 
(\ref{maux}), there exists a compact subset $K \subseteq I^-(p)$ such that $m(K)>m(I^-(p))-\epsilon =t^-(p)-\epsilon$ and, by inner continuity, $K \subset I^-(p_n)$ for large $n$. 
Thus, $t^-(p_n) \geq m(K) >t^-(p)-\epsilon$, as required. 

\smallskip

\noindent (b) $(\Rightarrow)$. Completely analogous to the previous case, taking now $K$ as a compact subset of $M\backslash \overline{I^-(p)}$ with $m(K) > m(M\backslash \overline{I^-(p)})-\epsilon$
and, then, for large $n$: $t^-(p_n) \leq m(M)-m(K) <t^-(p)+\epsilon$.

$(\Leftarrow )$. If $I^-$ is not outer continuous, there exists a compact $K\subset M\backslash \overline{I^-(p)}$ and a sequence $\{p_n\}\rightarrow p$ such that each $\overline{I^-(p_n)} \cap K$ contains at least one point $r_n$. Thus, $r_n\rightarrow r \in K$, up to a subsequence, and  
choose $s<<r$ in $M\backslash \overline{I^-(p)}$. As the chronological relation is open \cite[Lemma 14.3]{O}, there exist neighborhoods $U,V \subset M\backslash \overline{I^-(p)}$ of $s, r$, resp., such that $U \subset \cap_{p'\in V} I^-(p')$, and, thus, $U \subset I^-(p_n)$ for large $n$. Now, choose a sequence $\{q_j\}\rightarrow p$ satisfying
$$ p<< q_j << q_{j-1} , \quad \hbox{for all}\; j .$$
Then, $U \subset I^-(q_j)$ for all $j$ and, putting $\epsilon = m(U)>0$:
$$ t^-(q_j) = m (I^-(q_j)) \geq m(I^-(q)) + m(U) = t^-(q) + \epsilon ,$$
as required.
\cvd

\begin{prop} \label{p33a}
Volume function $t^-$ (resp. $ t^+$) is continuous if and only if the set valued function 
$I^-$ (resp. $I^+$) is outer continuous.
\end{prop}
Thus, recalling that causal continuity means the outer continuity of $I^\pm$ plus the property of being distinguishing, characterized in Proposition \ref{voldis}(b), we obtain finally:

\begin{teor} \label{caucon}
A spacetime $(M,g)$ is causally continuous if and only if the volume functions $t^-, t^+$ are (continuous) time functions.
\end{teor} 

\begin{rema} {\em (1) The continuity of the volume functions can be characterized in terms of reflectivity \cite{Di}, \cite[Proposition  3.21]{BEE}. As reflectivity is closely related to causal continuity by Hawking, Sachs' characterizations \cite{HS}, 
this   yields an alternative proof of Theorem \ref{caucon} (see also \cite[Theorem 3.35]{BEE}).

(2) Proposition \ref{voldis} and Theorem \ref{caucon} characterize causal properties of the spacetime by using  volume functions, in a way independent of the  chosen measure, whenever the conditions in Subsection \ref{s31} are fulfilled.
}\end{rema}

\subsection{Causal simplicity}\label{s34}

\begin{lema} \label{s4l}
If $J^+(q)$ (resp. $J^-(q)$) is closed for all $q\in M$ then $I^-$ (resp. $I^+$) is outer continuous.
\end{lema}
{\em Proof.} Assume by contradiction that there exists $p\in M$, a compact subset $K\subset M\backslash \overline{I^-(p)}$ and a sequence $\{p_n\}\rightarrow p$ with an associated sequence $\{r_n\} \subset K$ such that $r_n \in \overline{I^-(p_n)}$. Let $r$ be the limit (up to a subsequence) of $\{r_n\}$, and choose $s\in I^-(r) \cap M \backslash \overline{I^-(p)}$. Then, for large $n$, $r_n\in I^+(s)$ and $p_n\in I^+(s) \subset J^+(s)$. Thus, as $J^+(s)$ is closed, $p\in J^+(s)$, i.e., $s\in J^-(p) \subseteq \overline{I^-(p)}$, a contradiction.

\cvd

\noindent Therefore, in any causally simple spacetime functions $I^-, I^+$ are outer continuous and, from Theorem \ref{caucon} and the definitions of causal simplicity and causal continuity:
\begin{prop} \label{s4p}
Any causally simple spacetime is causally continuous. In particular, volume functions $t^-, t^+$ are time functions.
\end{prop}

\begin{rema} \label{r41}{\rm
(1) Proposition \ref{s4p} is applicable to globally hyperbolic spacetimes, because they are causally simple \cite[Prop. 3.16]{BEE}, \cite[p. 412]{O}. Nevertheless, for such spacetimes a direct proof of the outer continuity can be  given. In fact, assuming global hyperbolicity, the proof of Lemma \ref{s4l} can be carried out replacing the last sentence by the following. Choose a point $q>>p$ 
and, for large $n$, take a causal curve $\gamma_n$ starting at $s$, which crosses $p_n$ and ends at $q$. By global hyperbolicity, the causal limit curve $\gamma$ of the sequence  $\{\gamma_n\}$, not only exists but will cross $r$ and $p$ \cite[Corollary 3.32]{BEE}, a contradiction.

(2) As an example, take Lorentz-Minkowski $\L^2$ in lightlike coordinates with $\partial_u$ future-directed, see Fig. \ref{fig1}. Now, consider the open square $|u|, |v| <2$, and remove the negative $v$ semiaxis, $-2<  v \leq 0$. For the resulting spacetime $M$, $J^-(q)$ is closed for any $q$, but this is not the case for $J^+(q)$ (the spacetime is not causally simple). In fact, $I^+$ is outer continuous and $t^+$ is continuous, but this is not the case of $I^-$, $t^-$, as the sequence $\{(1,1/n)\}\rightarrow (1,0)$ shows.  
}
\end{rema}

\begin{figure}[ht] 
\centering
\includegraphics[width=12cm]{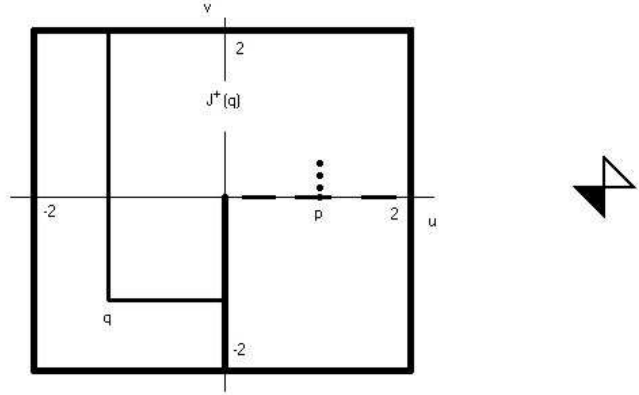}
\caption{$\L^2$, $g=-2dudv$, $M$ open subset (Remark \ref{r41}(2)), which is stably causal. 
$J^+(q)$ is not closed at $q=(-1,-1)$, and $t^-$ is not continuous at $p=(1,0)$.} \label{fig1}
\end{figure}

\section{Time functions and stable causality}\label{s5}

As we have seen, causally continuous spacetimes are those spacetimes such that $t^\pm$ are time functions. Now, one can consider the wider class of spacetimes which admit some time function, not necessarily $t^-$ or $t^+$. In the end, this class  turns out the class of stably causal spacetimes. Next, this equivalence will be revisited. 

Let Lor$(M)$ be the set of all the  Lorentzian metrics on $M$ (which can be assumed 
time-orientable  in what follows, without loss of generality). A partial (strict) ordering $<$ is defined in Lor$(M)$ naturally:
$ g<g' $ if and only if all the causal vectors for $g$ are timelike for $g'$. 

\begin{defi} \label{dsc}
A spacetime $(M,g)$ is stably causal if there exists $g'\in $Lor$(M)$ such that $g<g'$ and $g'$ is causal.
\end{defi}  
\begin{rema}\label{rsc}
{\em
(1) A stably causal spacetime can be understood as a causal spacetime which remains causal after opening slightly its lighcones. Notice that, if $g<g'$ then the metric $g_\lambda = g + \lambda (g'-g)$ is Lorentz for all $\lambda \in [0,1]$, and $0\leq \lambda_1 < \lambda_2 \leq 1$ implies $g_{\lambda_1}<g_{\lambda_2}$ (for each $v\in TM$, consider the line $\lambda \rightarrow g_\lambda(v,v)$ and apply $g<g'$). In particular, if $g'$ is causal then $g_\lambda$ is causal for all $\lambda\in [0,1]$. 

(2) An alternative definition is: {\em a spacetime $(M,g)$ is stably causal if and only if there is a fine $C^0$ neighborhood $U(g)$ of $g$ in Lor$(M)$ such that each $g_1 \in U(g)$ is causal}. Say,  ``stably causal'' means that the spacetime remains causal under $C^0$ fine perturbations. The $C^0$ topology is well-known (see for example \cite{HE} or \cite{Re}), and can be defined by using an auxiliary Riemannian metric $g_R$ as follows. For any $g\in $Lor$(M)$ and (positive) continuous function $\delta: M\rightarrow (0,\infty)$, let $U_\delta(g)=\{g_1\in \hbox{Lor}(M): |g-g_1|_R<\delta\}$, where $|g-g_1|_R$ denotes the pointwise norm induced by $g_R$ in the corresponding set of tensors. Now, a basis for the $C^0$-fine topology is defined as the set of all such $U_\delta(g)$ constructed for any $\delta$ and $g$. The independence of this topology with the choice of $g_R$ can be checked easily by taking into account that, for any other Riemannian metric $g'_R$, there exists positive continuous 
functions $\Lambda_1, \Lambda_2$ such that $\Lambda_1 g_R(v,v) \leq g'_R(v,v) \leq 
\Lambda_2 g_R(v,v)$ for all $v\in TM$ (choose $\Lambda_1$ at each point as the minimum eigenvalue of the endomorphism field associated to $g'_R$ by $g_R$, and $\Lambda_2$ as the maximum one). 
}\end{rema}
We will follow closely Hawking's \cite{Ha} (see also \cite[Prop. 6.4.9]{HE}) for the following result:

\begin{teor}\label{thaw}
Any stably causal spacetime admits a time function.
\end{teor}
{\em Proof.} By Remark \ref{rsc}(1), there exists a one-parameter family of metrics $g_\lambda$, $\lambda \in [0,2]$ which satisfies: (i) $g_0=g$, (ii) $g_\lambda$ is causal, for all $\lambda \in [0,2]$, and (iii) $\lambda < \lambda' \Rightarrow  g_\lambda < g_{\lambda'}$. Given $A\subset B\subseteq M$, the chronological past of $A$ relative to $B$  respect to $g_\lambda$ (those points reachable by a past-directed $g_\lambda$-timelike curve  starting at $A$ and enterely contained in $B$),  will be denoted as $I^-_\lambda(A;B)$. The measure will be normalized $m(M)=1$, and $t_\lambda^-$ will denote the past volume function respect to $g_\lambda$; recall $0< t_\lambda^-(p) <  1, \forall p \in M$. 

The required time function will be:
$$t(p)= \int_0^1 t_\lambda^-(p) d\lambda .$$
In fact, $t$ is a generalized time function because so is each $t_\lambda^-$ (Subsection \ref{s32}). To prove upper semi-continuity, fix $p\in M$ and $\epsilon \in (0,1)$. Take an open neighborhood $\B$ of $p$ ($\B$ can be chosen a convex neighborhood, \cite[p. 129]{O}) with $m(\B)<\epsilon/2$.

\smallskip

\noindent {\em Claim.} There exists a neighbourhood $V$ of $p$ such that:
\be \label{eclaim}
I^-_\lambda(V;\bb) \cap \partial \B \subset I^-_{\lambda + \frac{\epsilon}{2}}(p;\bb) \cap \partial \B , \quad \forall \lambda \in [0,1].
\ee
Notice that (\ref{eclaim}) would imply
$$I^-_\lambda(q;M) \backslash  \bb \subset I^-_{\lambda + \frac{\epsilon}{2}}(p;M)  \backslash \bb , \quad \forall q \in V, \quad \forall\lambda \in [0,1] , $$
and, thus,
$$ t^-_\lambda(q)\leq t^-_{\lambda + \frac{\epsilon}{2}}(p) + \frac{\epsilon}{2}, 
\quad \quad \forall q \in V. $$
Therefore the upper semi-continuity would follow directly:
$$ t(q) \leq \int_0^1 t^-_{\lambda + \frac{\epsilon}{2}}(p) d\lambda +
\frac{\epsilon}{2} = t(p)- \int_0^{\epsilon /2} t^-_{\lambda}(p) d\lambda
+\int_1^{1+ \epsilon/2} t^-_{\lambda}(p) d\lambda + \frac{\epsilon}{2}$$ 
$$< t(p) + \epsilon , \quad \forall q\in V.
$$
The claimed neighborhood $V$, can be found because of: (a) fixed $0\leq \lambda < \lambda' \leq 2$, there exist a neighborhood $V[\lambda, \lambda']$ of $p$ such that 
\be \label{eclaimbis}
I^-_\lambda(V[\lambda, \lambda'];\bb) \cap \partial \B \subset I^-_{\lambda'}(p;\bb) \cap \partial \B , 
\ee
(of course $V[\lambda, \lambda']$ is not unique, and any neighborhood of $p$ included in $V[\lambda, \lambda']$ also satisfies (\ref{eclaimbis})), (b) if $\lambda_1 <\lambda_2 <\lambda_2'<\lambda_1'$ then $V[\lambda_2, \lambda_2']$ also satisfies (\ref{eclaimbis}) for $\lambda=\lambda_1, \lambda'=\lambda'_1$ (that is, $V[\lambda_2, \lambda_2']$ can be taken as 
$V[\lambda_1, \lambda_1']$), and (c) choose $n\geq 2/\epsilon$, the open neighborhood
$$V= \cap_{i=0}^{2n} V[\frac{i}{2n}, \frac{i+1}{2n}]$$ 
can be taken as $V[\lambda, \lambda']$ for any $\lambda, \lambda'$ with $\frac{1}{n} \leq \lambda'-\lambda$, $\lambda \in [0,1]$, in particular, for $\lambda' =\lambda +\epsilon/2$.

Finally, the lower semi-continuity can be obtained analogously by claiming the existence of a neighborhood $V$ such that: 
$$
I^-_\lambda(p;\bb) \cap \partial \B \subset I^-_{\lambda + \epsilon/2}(q;\bb) \cap \partial \B \; , 
\quad \forall q \in V, \quad \forall \lambda \in [0,1].
$$
\cvd

It is straightforward to check that, if a spacetime admits a time function, then it is causal (and also strongly causal). As pointed out by Hawking \cite{Ha},  \cite[Proposition 6.4.9]{HE}, the existence of a temporal function $\Tau$ implies stable causality, because the spacetime will remain causal under $C^0$-fine perturbations of the metric  such that the lightcones do not touch the hypersurfaces at constant $t$. Let us detail two formal arguments:
\begin{lema} \label{ltemsc}
If a spacetime admits a temporal function $\Tau$ then it is stably causal.
\end{lema}
{\em Proof 1}, \cite[Prop. 6.4.9]{HE} (by using Definition \ref{dsc}.) As causality is a conformal invariant,  assume $g(\nabla \Tau, \nabla \Tau)=-1$ without loss of generality.  Now, the metric can be written as 
$$ g= -d\Tau^2 + h$$
where $h$ is the restriction of $g$ to the bundle orthogonal to $\nabla \Tau$ (up to natural identifications). Then, consider the  one parameter family of metrics
$$ g_\lambda = -\lambda d\Tau^2 + h ,  \quad \quad \lambda >0 .$$
Clearly, $\Tau$ is still a temporal function for each $g_\lambda$. Thus, $g_\lambda$ is always causal,  and  $g=g_1 < g_2$, as required. 

\smallskip

\noindent {\em Proof 2}, \cite{Ha} (by using Remark \ref{rsc}(2).) Let $S$ denote any hypersurface at constant value of $\Tau$ through a generic $p\in M$. Consider the set $U(g)\subset $Lor$(M)$ defined as follows: $g'\in U(g)$ iff the $g'$-causal cones at each $p\in M$ do not touch the tangent space at $p$ of $S$. One can check that $U(g)$ is open (recall Remark \ref{rsc}(1) and the possibility to use partitions of the unity). Even more, each $g'\in U(g)$ is causal. In fact, $\Tau$ is strictly increasing on any future-directed causal curve $\gamma$ at any  $p$, because  $S$ remains spacelike for $g'$ and, then, is crossed by $\gamma$ in the direction where $\Tau$ increases. 

\cvd

Now, the equivalence between stable causality and the existence of time functions will be completed  by the following result (see Fig. \ref{fig2}).
\begin{teor} \label{timetem}
If a spacetime admits a time function $t$ then it admits a temporal function $\Tau$.
\end{teor}
Nevertheless, the proof of this result will be postponed until  the whole smoothing procedure in 
globally hyperbolic spacetimes be finished.

\begin{figure}[ht] 
\centering
\includegraphics[width=12cm]{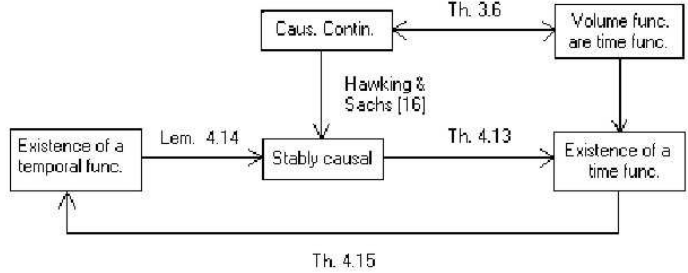}
\caption{Summary of implications} \label{fig2}
\end{figure}

\begin{rema} \label{reojo} {\em
As we have seen, Theorems \ref{caucon}, \ref{timetem} and Lemma \ref{ltemsc},  prove the implication ``causally continuous $\Rightarrow$ stably causal''. An alternative proof without time functions can be seen in \cite[Prop. 2.3]{HS}. 

Nevertheless, in order to prove the implication ``existence of a time function $\Rightarrow$ stably causal'', we do not know an alternative to Theorem \ref{timetem} and Lemma \ref{ltemsc}. In fact, a difficulty arises in the two proofs of Lemma \ref{ltemsc}, even when $\Tau$ is a smooth time function with lightlike gradient  at some point (say, for $\Tau(x,y)= y -$arctag$(x)$ in $\L^2$): the levels of $\Tau$ in $(M,g)$ do not remain achronal for all the metrics in a $C^0$ neighborhood of $g$. }
\end{rema}

\section{Geroch's topological splitting theorem} \label{s4}
Recall that function $t$ below is a time function by Remark \ref{r41}(1).
\begin{lema}\label{l516}
In a globally hiperbolic spacetime, the continuous function
\be \label{ft} t(p)=\log \left(-\frac{t^-(p)}{t^+(p)}\right) =\log \left(\frac{m(I^-(p))}{m(I^+(p))}\right)\ee
satisfies:
\be \label{elt}
\lim_{s\rightarrow a}t(\gamma(s))=-\infty,
\quad  \quad \lim_{s\rightarrow b}t(\gamma(s))=\infty
\ee
for any inextendible future-directed causal curve $\gamma: (a,b)\rightarrow M$. 
\end{lema}
{\em Proof.}
It is sufficient to check:
$$ \lim_{s\rightarrow a}t^-(\gamma(s))= 0, \quad  \lim_{s\rightarrow b}t^+(\gamma(s))=0.$$
Reasoning for the former, notice that, from (\ref{maux}), it is enough to show that, fixed any compact subset $K$,  then 
$K\cap I^-(\gamma(s_0)) =\emptyset $ for some $s_0\in (a,b)$ (and, thus, for any $s<s_0$). Even more, we can assume $K\subset I^+(q)$ for some $q\in M$ (otherwise,  write $K$ as the union $K=K_1 \cup \dots \cup K_l$ where each $K_j$ is compact and lies in some $I^+(q_j)$, and take  $s_0$ equal the minimum of the ones obtained for each $j$). Choose any point on the curve, $q= \gamma(c)$ for some $ c\in (a,b)$, and assume by contradiction the existence of a sequence $p_j=\gamma(s_j), s_j\rightarrow a, s_j\in (a,c)$, with an associate sequence $r_j \in K \cap I^-(p_j)$. Up to a subsequence, $\{r_j\}\rightarrow r$, and choosing $p<<r$, one has $p<<p_j<<q$, and $\gamma|_{(a,c]}$ lies in the compact subset $J^+(p)\cap J^-(q)$. That is, $\gamma$ is totally imprisoned to the past, an obvious contradiction with strong causality (see for example \cite[Prop. 6.4.7]{HE}). \cvd

\begin{teor} \label{tger}
Assume that the spacetime $M$ is globally hyperbolic. Then, function  $t: M \rightarrow \R$ in (\ref{ft}) is  an onto Cauchy time function.

Even more, given such $t$ and fixed a level $S_0= t^{-1}(0)$, a homeomorphism can be constructed
\begin{equation} \label{ed1}
\Psi: M\rightarrow \R \times S_0 , \quad z \rightarrow (t(z), \rho(z)) .
\end{equation}

\end{teor}
{\em Proof}. As $t$ is a time function, each level  $S_c$ is an acausal hypersurface. In order to check that any inextendible timelike curve $\gamma$ crosses $S_c$, recall that $\gamma$ can be reparametrized on all $\R$ with $t$, and (\ref{elt}) will also hold under any increasing continuous reparametrization of $\gamma$. Thus, assuming that this reparametrization has been carried out, $\gamma(c) \in S_c$.

For the last assertion, choose any smooth complete  timelike vector field $X$ (the completeness can be achieved dividing the vector field by its pointwise $g_R$-norm, where $g_R$ is any complete Riemannian metric), and define $\rho(z)$ as the unique point of $S_0$ crossed by the integral curve of $X$ through $z$.
\cvd

\begin{rema}\label{r411} {\em 
(1) Notice that even in a globally hyperbolic spacetime functions $t^-, t^+$ (and then $t$) 
may be non-differentiable (see Fig. \ref{fig3}) and, therefore, Theorem \ref{tger} yields a splitting only at a topological level. If $t$ were a  Cauchy temporal function then the vector field $X$ could be chosen pointwise proportional to  $\nabla t$, and an orthogonal splitting would be obtained.

(2) Geroch also proved that, if a spacetime $(M,g)$ admits a  Cauchy hypersurface $S$, then  the spacetime is globally hyperbolic (the proof is simpler, see \cite{Ge} or, for example, \cite[p. 422]{O}).  Then, Theorem \ref{tger} ensures that $M$ can be foliated by Cauchy hypersurfaces. Nevertheless, it does not ensure that $S$ is one of the leaves of the foliation. In fact, $S$ maybe non-acausal, and this cannot hold for the levels of a time function, Remark \ref{r22}. 
}\end{rema}

\begin{figure}[ht]
\centering
\includegraphics[width=12cm]{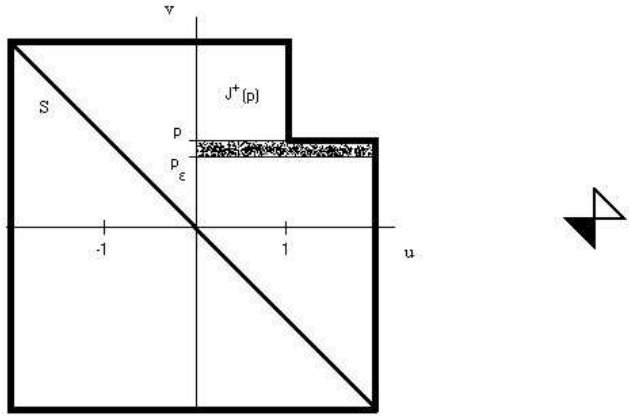}
\caption{$M \subset \L^2$, (coord. $u,v$). 
$M=\{(u,v)\in \L^2: |u|, |v| <2\}\backslash \{(u,v)\in \L^2: u, v \geq 1\}$; $p=(0,1), p_\epsilon=(0,1-\epsilon)$. Diagonal $S$ is a Cauchy hypersurface. For the natural $g$-measure, $t^+(p_\epsilon ) = 2 \epsilon + t^+(p)$ when $\epsilon>0$, and $t^+$  is not smooth.
} \label{fig3}
\end{figure}

\section{Orthogonal splitting and temporal functions}\label{s6}
Next, we will sketch how to strenghten  Geroch's splitting in an orthogonal one, and  time functions in temporal functions. We refer to the original articles \cite{BS1, BS2} for full proofs, or to \cite{BSc} for a more detailed summary.

In what follows, $(M,g)$ will be either a globally hyperbolic spacetime --Subsections \ref{s61}, \ref{s62}-- or a spacetime which admits a time function (and  will turn out stably causal, Remark \ref{reojo}) --Subsection \ref{s63}. Function $t$ will be, accordingly, either a Cauchy time function or just a time function. At each case, $S_t=t^{-1}(t)$ will be a topological hypersurface, either Cauchy or only acausal.
\subsection{Existence of a spacelike Cauchy hypersurface}\label{s61}

\begin{teor} \label{tsuave}
Any globally hyperbolic spacetime admits a smooth spacelike Cauchy hypersurface $S$.
\end{teor}
To prove it, we use the following result, which is a consequence of intersection theory:
\begin{lema}\label{pp}
Let $S_1, S_2$ be two Cauchy hypersurfaces of a globally hyperbolic spacetime with $S_1 << S_2 $  and $S$  a connected closed spacelike hypersurface.
 If $S_1 << S << S_2$ then $S$ is a Cauchy hypersurface.
\end{lema}
Thus, fixed two Cauchy hypersurfaces 
$S_1 << S_2$   as in Theorem \ref{tger} (writing $S_{t_i} \equiv S_i $; $ t_1<t_2$)
it is enough to find a hypersurface $S$ as in Lemma \ref{pp}. Such a $S$ can be obtained as the inverse image of any regular value $s\in (0,1/2)$ of any function $
h: M \rightarrow [0, \infty)
$
which satisfies:

\begin{itemize}
\item[1. ] $h(t,x)=0$, if $t\leq t_1$.
\item[2. ]  $h(t,x) > 1/2$, if $t = t_2$. 
\item[3. ] The gradient of $h$ is timelike and past-pointing on the open subset 
$V=h^{-1}((0,1/2)) \cap I^-(S_{2})$.
\end{itemize}
The construction of function $h$ is carried out in two closely related steps: the first one is the construction of a local function $h_p$ around each $p\in S_2$, and the second one the global definition of $h$ as a locally finite sum of such functions. The main difficulty is the following. In Riemannian Geometry, global objects are constructed frequently from local ones by using partitions of the unity. Nevertheless, for our problem, the causal character of the gradient of functions in the partition would be, in principle, uncontrolled. Then,  $h$ will be constructed by using  the paracompactness of $M$ but avoiding  partitions of the unity.

\smallskip

\noindent {\bf Step 1: constructing $h_p$.} Fix $p \in S_{2}$, and a convex neighborhood of $p$, ${\cal C}_p \subset
I^+(S_{1})$ (${\cal C}_p$ is a normal starshaped neighborhood of any of its points). 
Define $h_p$ as any smooth function with support in ${\cal C}_p$ which satisfies on ${\cal C}_p \cap J^-(S_2)$:
$$
h_p(q) = e^{d (p',p)^{-2}} \; \cdot \, e^{-d(p',q)^{-2}} , 
$$
for some $p'\in I^-(p) \cap {\cal C}_p$, where $d$ is the time-separation (Lorentzian distance) on ${\cal C}_p$. 
Notice that $h_p: M \rightarrow [0, \infty)$ satisfies:

(i) $h_p(p)=1$.

(ii) The support of $h_p$ (i.e., the closure of $h_p^{-1}(0, \infty)$)  is compact and included in ${\cal C}_p$.

(iii) If $q \in J^-(S_{2}) $ and $h_p(q) \neq 0$ then $\nabla h_p(q)$ is timelike and past-pointing.

\smallskip

\noindent {\bf Step 2: global function $h$.} The open subsets $W_p= h_p^{-1}(1/2, \infty ),$
$p\in S_2$, cover $S_2$. By the paracompactness of $M$ (and assuming that the ${\cal C}_p$'s are small, say, of diameter smaller than 1 for some auxiliary complete Riemannian metric), one can find a numerable locally finite subset of the $W_p$'s which cover $S_2$. Then, the sum of the corresponding functions $h_p$ is the required $h$.

\begin{rema} \label{rhabil} {\em 
If $t:M\rightarrow \R$ were a time function, not necessarily Cauchy, then the same procedure yields a closed connected acausal spacelike hypersurface $S$ between any two levels of $t$, $S_1<<S_2$.
}
\end{rema}
\subsection{Orthogonal splitting for globally hyperbolic spacetimes}\label{s62}
\begin{teor}
Any globally hyperbolic spacetime $(M,g)$ admits a Cauchy temporal function and, thus, it is isometric to the smooth product manifold
$$
\R \times {\cal S}, \quad \langle \cdot , \cdot \rangle = - \beta\,d\Tau^2 + \bar g_\Tau 
$$
where 
$\beta:\R \times {\cal S} \rightarrow (0,\infty)$ is a smooth function, 
$\Tau: \R\times {\cal S} \rightarrow \R$ the natural projection, each  ${\cal S}_\Tau$ (slice at constant $\Tau$)   a spacelike Cauchy hypersurface, and $\bar g_\Tau$ a Riemannian metric on each $S_\Tau$, which varies smoothly with $\Tau$.
\end{teor}
By taking into account Remark \ref{r411}(1), we have to prove only the existence of a Cauchy temporal function, which will be denoted also by $\Tau$. Notice that the spacelike Cauchy hypersurface $S$ in Subsection \ref{s62}, was found between $S_1, S_2$, as a regular value of $h$. The proof will be carried out in three steps: (I) to check that $S_2$ or, say, any Geroch's $S_t$, can be covered by spacelike Cauchy hypersurfaces obtained as regular values of some function  $\hat \tau_t$ around $S_t$, (II) to check that this also holds for a rectangular neighborhood $(-\epsilon,\epsilon)\times S_t$ of each $S_t$, by constructing an appropiate ``temporal step function'' $\tau_t$, and (III) to obtain $\Tau$ as an appropiate sum of some of such $\tau_t$'s. 

\smallskip

\noindent {\bf Step I.} The aim is to prove that, for each Geroch's $S_t$, there exists a smooth function $\hat \tau_t: M\rightarrow \R$} which satisfies:
\ben
\item $\nabla \hat \tau_t$ is past timelike in $V_t :={\rm Int(Supp}(\nabla \tau_t))$.

\item $-1 \leq \hat \tau_t \leq 1$.

\item $\hat \tau_t(J^+(S_{t+2})) \equiv 1$, $\hat \tau_t(J^-(S_{t-2})) \equiv -1$. 

\item $S_{t} \subset V_t $ ($\nabla \hat \tau_t$ does not vanish on $S_t$).
\end{enumerate}

\noi This function can be obtained as the combination
$$\hat \tau_t = 2 \; \frac{h^+}{h^+ - h^-} -1$$
where $h^+ \geq 0$ is, essentially, function $h$ constructed in Subsection \ref{s62} 
for $S_t=S_{t_2}$, and $h^- \leq 0$ is constructed similarly, but with modified technical properties.

\smallskip

\noindent {\bf Step II.} The fourth requirement on previous function $\hat \tau$ can be strengthened to obtain function $\tau_t: M\rightarrow \R$ ({\em temporal step function around $S_t$})
which satisfies the three first items plus:
\begin{itemize}
\item[4'.] $S_{t'} \subset V_t $, for all $t' \in (t-1,t+1)$.
\end{itemize}

\smallskip

\noi The idea is to check first that, for any compact $K \subset  [t-1,t+1]\times S$,  
previous function $\hat \tau_t$ can be chosen such that $\nabla \hat \tau_t \neq 0$ on $K$. 
Then, choose an increasing sequence of compact subsets 
$\{ K_j \}$, with  $[t-1,t+1] \times S \subset \cup_jK_j \subset 
(t-2,t+2) \times S$, and construct one such function
$\hat \tau_t[j]$ for each $K_j$. Then, define
$\tau_t= \sum_j \hat \tau_t[j]/C_j$, where the $C_j$'s are positive constants chosen to make the series and its derivatives uniformly convergent and, then, $\tau_t$ smooth.

\smallskip
\noi {\bf Step III.} Required $\Tau$ is a sum of temporal step functions:

$$ \Tau = \tau_{0} + \sum_{k=1}^{\infty} (\tau_{-k} + \tau_k )$$ 
In fact, $\Tau$ is temporal, because subsets $V_{t=k}, k\in \Z$ cover all $M$ (and the timelike cones are convex). And the levels of $\Tau$ are Cauchy  because, for each inextendible timelike curve $\gamma: (a,b) \rightarrow M$, the limits in formula (\ref{elt}) hold for $\Tau$ in the role of $t$.

\subsection{Temporal functions in stably causal spacetimes}\label{s63}
Finally, let us sketch the proof of Theorem \ref{timetem} (essentially, a simplification of the steps to construct the Cauchy time function above). Let $t$ be a time function, choose $p \in M$ and let $S$ be the level hypersurface of $t$ through $p$. Then, $S$ is closed, achronal and separates $M$ ($M\backslash S$ 
 is the disjoint union of two non-empty open subsets). Now:

(i) For each $p\in M$, there exists a function $\tau_p$, $-1\leq \tau_p \leq 1$ such that  $\nabla \tau_p$ is: (a) either timelike or 0 everywhere, and (b) timelike on a neighborhood of $p$ (in fact, of $S$). Essentially, this function is constructed as  $\hat \tau_t$ in Step I (recall Remark \ref{rhabil}).

(ii) Given any compact subset $K \subset M$, a similar function $\tau$, which satisfies not only (a) but also (b) for all $p \in K$,  can be obtained as a finite sum of functions constructed in (i).

(iii) Choosing a sequence of nested compact  subsets $K_j$ which cover $M$ (as in equation (\ref{succom})), taking the   corresponding function $\tau_j$ obtained in (ii), and summing a series 
$\Tau = \sum_j \hat \tau_j/C_j$
(with constants $C_j>0$ which make smooth the sum of the series), the required $\Tau$ is obtained.  

\begin{rema}
{\em Notice that, fixed the closed acausal hypersurface $S$, its Cauchy development $D(S)$ (with the conventions in \cite{O}) is an open globally hyperbolic subset, and any acausal Cauchy hypersurface  $R$ of $D(S)$, is also acausal for $M$. Even more, 
$D(R)=D(S)$, $\partial D(R) \cap I^\pm(R)= \partial D(S) \cap I^\pm(S)$; thus,
$R$ will be also closed in $M$, and will separate $M$. As a consequence, any temporal step function on $D(S)$ can be extended trivially by $\pm 1$ to  all $M$ (this yields an alternative way to obtain the function $\tau_p$ required in (i)).
}
\end{rema}
\section{Final discussion}\label{s7}

According to Arnold \cite{Ar}, open problems maybe formulated in the Russian or the French styles. The former is to mention the simplest non-trivial case; the latter, to wonder the  most general question making impossible further generalizations. Sachs and Wu  \cite[p. 1155]{SW} followed the Russian style: they only asked for the existence of a smooth Cauchy hypersurface (a case of obvious interest, because Einstein's equations are naturally posed on smooth -even spacelike- Cauchy hypersurfaces). When we obtained the full orthogonal splitting  plus the existence of temporal functions \cite{BS2},  we thought to have answered the French problem. Say, from the conceptual viewpoint it seems satisfactory, and that  is what one needs for applications as 
Morse Theory \cite{Uh}, quantization \cite{Fu},  variational methods \cite{Ma} or splittings of stably causal spacetimes \cite{GK}. But, well, speaking with our colleagues, and reflecting on our own results, the author noticed that it is impossible  to ensure that a problem has been formulated in a ``truly French'' style --otherwise, the question ``find related developments'' remains open. The following three questions arise naturally from our solution, and may be of interest in its own right:

\ben
\item Given a causally continuous spacetime, volume functions $t^-, t^+$ are (continuous) time functions for any admissible measure. It is easy to find examples such that volume functions are not smooth for {\em some} admissible measure $m$ (this happens in the globally hyperbolic example of Fig. \ref{fig3}). Now, one can wonder:  {\em in this case,  is it possible to find another admissible measure such that $t^-, t^+$ become smooth and even temporal?} 

Notice that a positive answer would yield, in particular, an alternative way to the smoothing procedure for globally hyperbolic spacetimes in Section \ref{s6}. Nevertheless, even in this case the existence of a temporal function in  stably causal spacetimes would need a new procedure (as in Subsection \ref{s63}), because, in the  non-causally continuous case,  volume functions are never continuous.

\item Intuitively, the spacelike Cauchy hypersurface obtained in Subsection \ref{s61} can be seen as a ``smoothing'' of $S_2$ in the following sense. If $t_1<t_2$ then the spacelike Cauchy hypersurface ${\cal S}$ is contained in    $t^{-1}(t_1, t_2)$. Thus, ${\cal S}$ approaches $S_{2}$ as $t_1$ approaches $t_2$, . But the whole procedure to obtain the (Cauchy) temporal function does not have such a direct interpretation, and the final temporal  function $\Tau$ maybe very different to the original time one $t$. Nevertheless, probably one can keep track of the different steps, in order to answer: {\em can the  topological elements (say, a Cauchy time function) be approximated in a natural topology by smooth elements (Cauchy   temporal functions)?}

\item Given a Cauchy hypersurface $S$, as the spacetime is globally hyperbolic, there exists a function whose levels foliate the manifold by means of Cauchy hypersurfaces. Nevertheless, these hypersurfaces are rather unrelated to the original one, even at the topological level of Geroch's result, Remark \ref{r411}(2). Thus, we can wonder: can the whole spacetime be foliated by Cauchy hypersurfaces, being one of the leaves the original one $S$? Even more, in the case that $S$ is additionally, (a) acausal (recall Remark \ref{r22}), (b) acausal and smooth, or (c)  spacelike: {\em is there a function $t$ such that $S$ is one of its levels and $t$ is (a) a Cauchy time function, (b) a smooth Cauchy time function, or (c) a  Cauchy temporal function?}

\een

\section*{Acknowledgments}

The author acknowledges comments by Prof. P.E. Ehrlich, N. Ginoux and, especially, A. N. Bernal. Partially supported by the MCyT-FEDER Grant MTM2004-04934-C04-01.

  \end{document}